# A Frugal AI Framework for India

## Responsible and Right-sized AI

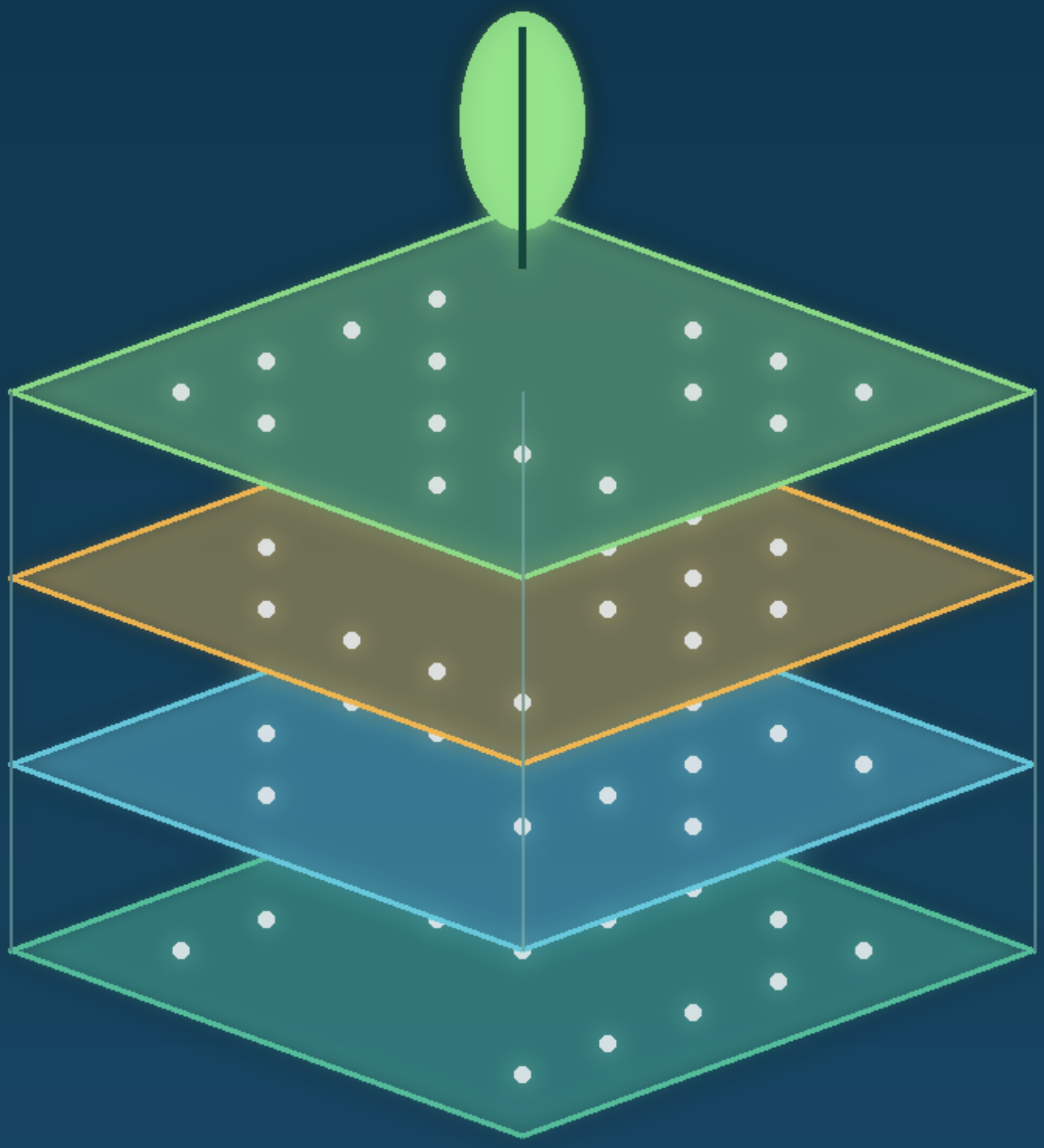

Centre for Responsible AI (CeRAI)
Wadhwani School of Data Science and AI (WSAI)
Indian Institute of Technology Madras

CeRAI
Centre for Responsible AI

WSAI IITM

# A Frugal AI Framework for India

## Responsible and Right-sized AI

## Contributors

**Prof. Balaraman Ravindran**
Head, Wadhwani School of Data Science and AI (WSAI) and Centre for Responsible AI (CeRAI), IIT Madras

**Mr. K. S. Venkataraghavan**
Senior Advisor, CeRAI, IIT Madras

**Dr. Gokul S Krishnan**
Research Director, CeRAI, IIT Madras

**Prof. Balaji Srinivasan**
Professor, WSAI, IIT Madras

**Prof. Balkrishna C. Rao**
Professor, Department of Engineering Design, IIT Madras
Adjunct Professor, Institute of Machine Design and Tribology (IMKT), Leibniz University Hannover

*The contributors acknowledge the use of Claude (Anthropic) and Perplexity for assistance in the preparation of this paper. All the content was reviewed, verified, and edited by the authors, who take full responsibility for the accuracy and integrity of the published work.*

# Contents

# Executive Summary

India's AI debate often swings between two insufficient views: replicating the compute-intensive, capital-heavy foundation-model programmes similar to that of the United States or China, or remaining a downstream adopter of foreign AI with localised interfaces and safeguards. Neither serves a country that must deliver AI to a very large, diverse population across uneven infrastructure and within constrained budgets.

The Centre for Responsible AI (CeRAI), IIT Madras, proposes a third path: **Responsible Frugal AI**, which applies principle-based frugal engineering alongside Responsible AI. It supports Green AI by cutting compute, hardware and energy use, and Responsible AI because smaller, better-scoped systems are easier to govern. Responsible Frugal AI is right-sized AI. It meets a fixed functional requirement with minimal resources, while keeping a verified margin, robustness and responsible governance. In practice, functional requirements are seldom fixed against the user's task, which is why many AI systems are over-specified or deployed without assessing their total cost of ownership. Responsible Frugal AI fixes the requirement first, asks whether AI is needed at all, and then chooses the smallest sufficient model to fulfil the task requirements. It builds on advanced frugal innovation and frugal engineering, disciplines with strong Indian roots, and is consistent with UNESCO's 2021 Recommendation on the Ethics of AI.

CeRAI applies frugality across four dimensions (data, compute, model and operations) through seven levers (data foundation, architecture choice, training efficiency, compression, adaptation, retrieval grounding, and serving and silicon) that span the AI lifecycle, each with practices and measures. The framework calls for public AI rails, vernacular-first design, measured resource use, human oversight for high-stakes decisions, and a cap on total resource consumption, so that efficiency gains are not cancelled out by a rise in overall AI usage.

India is resource-imbalanced rather than resource-poor: compute, connectivity, skills and capital exist but are unevenly distributed, and access to compute is the major constraint. Frugal principles offer the most viable route to population-scale deployment and fit naturally with India's digital public infrastructure. Sovereignty follows because India retains the capacity to specify, design, deploy, govern, operate, inspect and adapt right-sized AI systems, with Responsible AI principles such as accountability and transparency underpinning that autonomy.

A seven-point adoption checklist sets minimum conditions for a frugal deployment, from establishing the need for AI and declaring constraints to checking cost and data jurisdiction before scale-up. Eleven policy statements set out the commitments this white paper proposes, including frugality as the default for public-interest AI, shared public AI rails instead of duplicated stacks, and alignment with the India AI Governance Guidelines (2025). Together, they position AI in India as a responsible, frugal public-value system and as the route to maximising social value per rupee and per joule.

# I. Frugal Innovation, Frugal Engineering, Frugal AI

"Frugal " and " frugality" are well-known terms for being thrifty or economical. However, these terms have undergone massive reconceptualisation[1].

The idea of "innovation in a sandbox" to create an impossibly low-cost, high-quality new business model was practised by cultivating constraints[2].

"Frugal Engineering" grew in contrast to product development processes that became hardwired, and whose standard operating procedures worked against frugality[3].

Indian firms have notably developed an engineering discipline that minimises resource use and cost while satisfying a fixed functional requirement, guided by rigorous scientific and design principles. This discipline is commonly termed frugal or constraint-based innovation. Where such designs subsequently diffuse to advanced economies, the phenomenon is referred to as reverse innovation[4].

Frugal innovation extends efficiency principles by reframing the problem. Instead of simply improving resource use within given requirements, it first asks whether those requirements are nominal—that is, necessary and sufficient. It then seeks to meet them with the least resource use and cost.

Frugal innovations in turn are classified into Grassroots (GFIs) and Advanced Frugal Innovations (AFIs). GFIs are makeshift products/contraptions assembled from a mixture of parts without any formal application of science. AFIs, in contrast, engineer frugality through formal, methodical frugal engineering with cutting-edge science and technology. AFIs have appeared in sectors like astronomy in addition to aerospace, automotive, healthcare and others[5]. Hence, Frugal AI is an AFI that seeks to achieve low-resource, low-cost and high-quality simultaneously without trading off one against the other.

Frugality in resources and cost, while upping quality, can be seen as a principle of advanced frugal innovation for designing systems with minimal resource use and sustainability. This is always bounded by a defined threshold and a verified safety margin. It also applies to constrained, multi-objective optimisation, where functionality is the constraint and resources, cost, and robustness are the variables. This threshold is established from rigorous user needs and requirements, with an appropriate safety margin[5, 6].

It may also be noted that, as a fallout of economic developments in the early 2020s, such as the COVID-19 pandemic, ongoing geopolitical uncertainties, energy crises, and the global cost-of-living crisis, an emergent tendency to adopt simplicity and frugality voluntarily is increasingly apparent, both by choice and as a marked characteristic of social responsibility[5, 7, 8].

AI engineered with a principle-based approach such as frugality delivers the functionality users need under expected operating conditions while minimising required resources and costs[9].

In keeping with the AFI concept, Frugal AI does not trade engineering rigour for economy. It maintains a verified operating margin and the robustness or adaptability needed to prevent premature failure as conditions change[9].

The following also becomes more significant when considering what "Frugal AI" stands for.

- UNESCO Recommendation on the Ethics of Artificial Intelligence, adopted in 2021, states in Paragraph 86[10], "*Member States should ensure that AI actors, in line with the principle of proportionality, favour data, energy, and resource-efficient AI methods*."
- UNESCO's TVETipedia glossary includes the following[11]
    - Core Idea: Frugal AI is a paradigm shift towards designing AI systems that achieve high impact with minimal resources. It's about "doing more with less" across compute, energy, data, and capital.

- Key Principles: Frugal AI is guided by key principles: resource efficiency, sustainability, accessibility & inclusion, and impact & scalability. These ensure responsible and equitable AI development.

The term " Frugal " in Frugal AI thus has many connotations, and the following are noteworthy and considered significant in this framework, given the “Prime Contribution from IIT Madras.,

- IIT Madras’s contribution, which is continuously based on impactful innovation over imposed restraint; in short, this is Advanced Frugal innovation where products and services shall be designed for resource-constrained environments[2–4]
    - This considers environmental impact and use and shall be rooted in rigorous science, engineering design and innovation management[12–16].

# II. Frugal AI: A Broader Responsible AI Strategy for India

As noted above, Frugal AI has evolved from the principle-based approach of Frugal engineering itself[5] and initiatives such as “Green AI”[17] primarily focus on reducing environmental costs, including energy use and carbon footprint. These initiatives have benefited from measuring computational costs and improving efficiency alongside performance and environmental footprint [17–19].

Some commercial initiatives are focused on integrating artificial intelligence with clean technology to drive sustainability and economic growth[20]. A case study also reported critical success factors (CSFs) for AI implementation in sustainable frugal innovation and implementation[21].

At CeRAI, IIT Madras[22, 23]. “Responsible AI” includes safety, trustworthiness, fairness, privacy, transparency, accountability and governance. Frugal AI, along with “Responsible AI”, minimises harmful behaviour in data, model design, deployment, operation and governance. Its value-add comes from being grounded in a well-established, principled engineering approach that offers rigorous, evidence-based science and adequacy under constraints[10, 15, 24].

A Frugal AI system contributes to “Green AI” outcomes because it provides evidence-based pathways to use less compute, hardware, and energy. It supports “Responsible AI” because any smaller, narrower, and better-scoped systems are often easier to sustain, interpret and govern.

It can also be argued that these relationships are not automatic, because any system can be efficient yet unjust, or locally deployed yet environmentally wasteful. That is why CeRAI considers Frugal AI as a key part of a “Broader Responsible AI strategy for India ”[i.e. Responsible Frugal AI], not just a substitute for “Green AI".

Although many Frugal AI-related narratives evolved in parallel, this positioning is CeRAI IIT Madras’s India-specific contribution, aligned with both UNESCO’s 2021 Recommendation[10] and its glossary definition of Frugal AI[11].

# III. Why Frugal AI, why now?

Robustness has long been a key AI engineering optimisation criterion. Frugal AI thus requires combining minimal resource consumption, good functionality, robust operation under unplanned conditions, recovery from failure under nominal conditions, and low cost. With Frugal AI, performance and robustness become a declared threshold to meet with a verified margin, not just a ‘maximum factor’ to pursue. Frugal AI is not unique to India; nations have increasingly adopted this approach to AI[25].

## i. AI Performance and Frugal AI

Historically, AI engineering has treated performance as a principal design goal. Market requirements, however, had seldom considered cost-effectiveness, energy use, or environmental impact. Only as AI products faced rising costs and increasing sustainability expectations did performance come to mean a balance of capability, energy efficiency, and economic viability. ML research rewarded accuracy, bought with compute (*Refer to "Green AI" vs "Red AI"*); this was becoming expensive[17].

Thus, cost, energy use, and environmental impact entered the specification late, as sustainability expectations rose alongside inference bills [17–19, 26–28].

Performance improvements, though often invisible to users, depended on optimisation throughout the AI product lifecycle. It was reported that AI's impact, especially in downstream markets, is context-dependent and depends on enabling "inputs" such as data/datasets and cloud infrastructure/compute[29]. The following observations are especially relevant in this context.

- Performance was treated as a technical hurdle rather than a feature to engineer[30]
    - "For decades, programmers have been able to prioritise making the writing of code easier rather than making the code itself run faster," and that prioritising productivity over performance "has led to problematic strategies."[30]
- It was shown that with the new architectures, engineers can easily envision future energy-efficient systems consisting of a great many accelerators executing specific operations or algorithms, their interactions orchestrated to perform larger tasks and turned on and off as needed[31].
- It was reported that research ideas succeeded depending on whatever hardware happened to be available, fit with the software, not on scientific merit, essentially indicating "Hardware Lottery", i.e., the performance of an algorithm is fundamentally intertwined with the hardware and software it runs on[32].
- Progress has relied heavily on adding compute rather than engineering efficiency[33]
- Positive cases that efficiency had to be co-designed across algorithm and hardware rather than added afterwards were reported[34]
- Large firms' own scientific output from their premises had declined steadily, while their patents still drew on recent science. Increasingly, they relied on knowledge from universities and smaller firms and overlooked the scientific foundations and missed opportunities from advances in science and materials
- It was also reported that scaling compute, parameters, and data lowers loss predictably, yet some abilities appear abruptly beyond thresholds, such as chain-of-thought near $10^{23}$ FLOPs[35].
- Some studies observe that large language models appear to acquire certain capabilities abruptly as scale increases. However, it notes that continuous evaluation metrics suggest this discontinuity may be partly a measurement artefact. Because capability trajectories remain difficult to forecast, the explainer argues that policymakers require improved measurement and forecasting methodologies[36].
- Proponents cite nonlinear dynamics[37], a theoretical account of why AI capabilities can appear abruptly. Smooth increases in scale, data or training time can produce qualitative shifts once a system crosses a sensitive threshold.
- Proponents of emergence frequently cite grokking[38], a phenomenon of delayed generalisation in which a small transformer trained on a synthetic algorithmic task initially memorises its training data and, after extended optimisation, undergoes an abrupt transition to generalisation. Such models share the architecture of contemporary large language models but are reduced to one or two layers and approximately $10^5$ parameters (*Parameter counts for (current) frontier models are mostly undisclosed*), several orders of magnitude below frontier systems[39]. They are trained on programmatically generated data governed by a deterministic rule, such as modular arithmetic, for which every input admits a single, computable correct output. This controlled setting permits an unambiguous distinction between memorisation and rule acquisition and renders the network's internal computation fully tractable to mechanistic analysis. Within the emergence debate, grokking therefore serves as a principal small-scale demonstration that sudden capability gains can reflect genuine internal reorganisation rather than artefacts of metric choice[40].

- It was also shown that ReLU(Rectified Linear Unit -a popular activation function) MLP (Multilayer Perceptron) (simple feed-forward networks with max(0,x) neurons) trained on modular addition show latent behaviour during memorisation, even without generalising; this suggests grokking refines latent structure rather than discovering it[41].
- Recent surveys have reviewed LLM emergent abilities, critiquing definitions, contesting metric-mirage claims, linking emergence to loss thresholds, task complexity, quantisation and internal mechanisms, covering reasoning models and harmful emergent behaviours[42, 43].
- Several analyses argue that if emergent abilities arise from nonlinear training dynamics, and so cannot be reliably extrapolated from smaller training runs, then neither the capabilities of a larger model nor the risks linked to emergence can always be predicted from smaller-scale experiments[44–46]. If that is right, models should be evaluated in stages, with graded testing, auditing and continuous monitoring at each step up in scale, and evaluation deserves more investment. The argument does not, however, settle whether more compute or larger models are justified. Any conclusion on that question is an interpretation adopted in this document, not a finding of the cited studies.
- Some critiques have argued that if emergence is dynamical and not reliably extrapolable from smaller runs, capabilities, including emergence-related risks, cannot always be predicted from smaller-scale experiments. This strengthens the case for staged, graded evaluation, auditing and continuous monitoring at each scaling step. Such evidence supports greater investment in evaluation. It does not, by itself, establish whether larger computation or larger models are warranted. That inference is interpretive and not a finding of the cited literature[44–46].
- Engineering effort, i.e., AI engineering and technology processes, has tended to retrofit by prioritising continuity, scaling, and lifecycle optimisation, as well as customised generic hardware, over the disciplined integration of scientific advances to design better networks, more capable systems, and fit-for-purpose AI infrastructure [26]. AI organisations have likewise promoted continuous improvement of established products while moving more slowly to fundamentally new technologies[26, 47, 48].
- Compute has also grown much faster than memory bandwidth (the “memory wall”), so the field should work around this limitation through redesigns in model architecture, training, and deployment strategies[49].

However, it is now argued that this has led AI to be deployed in different directions.

1) Some AI systems are over-specified, offering more capability and complexity than the problem demands[50, 51].
2) Heavy investment goes into infrastructure and architecture before real-world testing; 56% of organisations skip a Total Cost of Ownership assessment before placing large workloads. This indirectly supports the point that capability is bought without checking cost fit[52].
3) Others are under-deployed, as in infrastructure monitoring, medicine, manufacturing and where capability exists but is not fitted to a cost-justified use[53–57].

These above observations follow from one omission, [i.e.] the functional requirement was seldom fixed against the user’s task, so there was neither anything to minimise against nor anything to verify under a safety margin.

Market forces clearly shaped this pattern. Advances in engineering, despite the gaps they left behind, have widened the range of solutions and opened new life cycles aimed at addressing performance shortfalls.

Much of the resulting innovation remained in research settings and bespoke deployments, operations within large organisations, rather than in broad, cost-optimal applications[58, 59].

Viewed critically from a socio-economic standpoint, many of these AI products were expensive because they originated in economies high in the knowledge chain, focused on fundamental innovation and design rather than manufacturing[60, 61].

As such, they appeared better suited to automation or to capital-intensive, rather than labour-intensive, work. With the emergence of globally distributed value and supply chains, the shift toward large-scale automation of manufacturing processes, and the rising cost of raw materials for hardware, the production, deployment, and operation of AI equipment and infrastructure thus became prime candidates for optimisation and cost rationalisation[62, 63].

Given the above, and to reiterate the necessity of Frugal AI, the following subsections [Subsection ii and Subsection iii] consider the aspects that have shaped AI discourse on Frugal AI, followed by discussions and policy deliberations related to various definitions of “Frontier AI”[64–68] and its associated significance[refer to subsection iv].

## ii. Is Bigger the Better?

Varoquaux et al. (2025) [24] show that increased scale is neither a sufficient explanation for performance improvements nor a prerequisite for all valuable applications, and that continued scaling carries disproportionate economic and environmental costs; it can also distort research priorities and concentrate decision-making power[24]. From this, we can draw some conclusions. Scaling need not always be the default, nor should it always be discouraged or argued against. This points to alternatives to *“serve to address the many pathological symptoms associated with the race to scale, from widening inequality to climate harms, at a common root*.” Though not explicitly mentioned, the following are some informed connections to “Frugal AI” that can be inferred from the work[24].

- **Value research on small systems**. Compare every large model against simpler baselines, and value work on uncertainty quantification, causality and calibration done on small models.
- **Report size and cost**. Disclose compute, energy and memory for both training and inference (for example, with Code Carbon), using metrics such as samples per kWh or throughput.
- **Holding reasonable expectations**. Don’t ask to run additional experiments more costly than those already performed and recognise that solving a problem that only appears at small scale is valuable.
- **Use a Pareto framing (**Refer to the Figure ‘Pareto Optimality and different contributions' as seen below[24]).

  Treat the state of the art as a Pareto frontier in the trade-off space between task performance and computing resources, i.e. performance against resources and credit resource-efficient progress (same performance with less compute). Also, consider resource-intensive progress (more performance from more compute). i.e. increasing task performance by increasing computing resources. We then see resource-efficient progress that decreases computing resources for the same task performance.

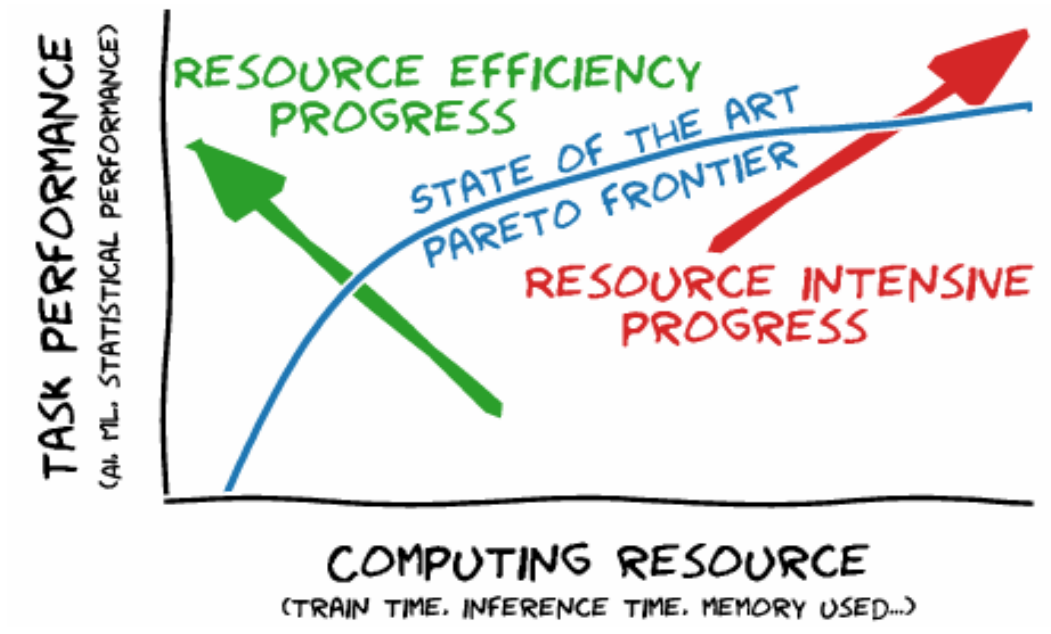

*Figure 1*

*Pareto Optimality and different contributions*

Figure 1 illustrates that state-of-the-art performance is best seen as a Pareto frontier in the trade-off space between task performance and computational cost. The contributions that get the most attention are usually resource-intensive progress, i.e. higher task performance bought with more compute. Resource-efficient progress deserves equal recognition, i.e. reaching the same task performance with less compute.

### iii. Model Distillation, Advancements in RL, Post-training Alignment, Reasoning Behaviour; RL and the Frugal AI Connection

(Model) Distillation (aka "Knowledge distillation") introduced ways to compress cumbersome models into forms that can be deployed widely, and it provided a pathway from capability developed through large-scale reinforcement learning to compact deployment models[69–72]. It was also shown that distillation could widen access to capable reasoning at substantially lower per-query energy, i.e. inference cost; however, capability improvements still depended on the large teachers themselves[73].

Post-training refers to procedures that adapt a pretrained base model's behaviour to human preferences, safety constraints, and, increasingly, specific task capabilities. Its modern form originates in reinforcement learning from human feedback (RLHF)[74]. With this approach, human comparisons between candidate outputs train a reward model, which in turn guides policy optimisation, indicating the significance of 'human preferences'[74–76]. These approaches were consolidated into a three-stage pipeline of supervised fine-tuning, reward modelling and reinforcement learning[76]. It is also reported that a new algorithm, "Direct Preference Optimisation" (DPO)[77], which the authors describe as computationally lightweight, solved the RLHF objective with a simple classification loss and no sampling during fine-tuning. More recent practice has shifted toward hybrid pipelines that combine supervised learning, direct preference optimisation [DPO], and reinforcement learning against verifiable rewards[78], often using memory-efficient policy-optimisation methods such as Group Relative Policy Optimisation (GRPO)[79, 80].

A growing body of work on reasoning behaviour suggests that reinforcement learning is largely bounded by the pretrained model's latent capacity. Reinforcement learning with verifiable rewards [RLVR][79, 81–83] improves single-attempt accuracy, yet base models match or exceed the trained models when many attempts are permitted. Some models even improve under random rewards, although this effect appears to be specific to model families[79, 81, 82].

Such behaviours, often attributed to reinforcement learning, such as the so-called "Aha moment"[84], are already present in base models before any RL is applied[84], and researchers have since analysed their characteristics in detail[85]. Taken together, these findings suggest that reinforcement learning mainly elicits and sharpens capabilities acquired during pretraining rather than creating new ones and that pretraining characteristics themselves have influenced RL performance[85].

As argued before, (Model) Distillation techniques[69–72] offer a complementary route. Whereas reinforcement learning appears constrained by its base model, distillation can transfer new reasoning patterns from teacher to student and extend the student's capabilities[83]. This allows reasoning developed through large-scale reinforcement learning in frontier models to carry over to smaller models suited to deployment[73], at a substantially lower inference cost[50, 71]. It is also reported that agents that learn directly from experience may shift the main bottleneck in capability development from data curation to reward curation[86].

From the above, we can note that

- RL's gains are bounded by pretraining.
- It follows that the expensive capability work is concentrated in a few large teacher models.
- Also, model distillation could become the frugal measure that can spread the above capability into small, cheap, deployable models[50, 69, 72].

Interestingly, RL connects to Frugal AI in two opposite ways. Historically, it is the least frugal paradigm in ML, because it needs huge numbers of samples and environment interactions. Yet it is also the most natural tool for frugality, because it can optimise resource use as an explicit objective. A frugal-AI programme should therefore consider RL both as something to make frugal and as a means of making other systems frugal.

### iv. AI Regulations and Frontier AI, Frontier Models, task-solving traces in their recent reasoning advances

There is no single agreed definition of "Frontier AI". Popular definitions use different criteria: "relative capability", "a capability plus risk-based classification", or "training compute"; some even call it "hype", and some combine model capability, size and FLOPS.

- **Relative Capability**: This defines Frontier AI as highly capable general-purpose AI models that can perform a wide variety of tasks and match or exceed the capabilities present in today's most advanced models[64]
- **Capability plus Risk-based classification**: In this classification, Frontier AI is defined as "Highly capable foundation models that could possess dangerous capabilities sufficient to pose severe risks to public safety"[66, 87, 88].
- **Statutory compute threshold /Training Compute**: Governments use a FLOPS threshold to determine how much compute was used to train models and to impose requirements based on that compute.[68, 89].
- **"Hype":** Frontier AI" is AI hype. It is hypothetical, not real. It distracts focus away from AI's actual harms by focusing on so-called "existential risk."[67].
- **Model Capability, Size and FLOPS**: This argues that setting "training compute" as a (fixed compute) threshold depends on whether a model's risks are tied to its training scale and its position relative to the frontier and concludes that this is based on the capability of the model as well as on size or FLOPS[65].

The capability-based approach makes the frontier a moving capability boundary rather than a fixed model size or computing threshold.

The risk-based approach suggests that linking regulatory thresholds to the largest training runs may keep the number of regulated models manageable.

The hype-based approach rests on the argument that these harms are real rather than theoretical, whereas the term "frontier AI" shifts our collective focus away from actual harms and toward hypothetical (doomsday) scenarios.

Also, the Frontier AI definition, adopted in the Frontier AI Safety Commitments at the AI Seoul Summit (2024)[90], refers to Frontier AI as highly capable general-purpose AI models or systems that perform a wide variety of tasks and match or exceed the capabilities of the most advanced models.

Current advances in frontier model capabilities stem from post-training fine-tuning using RL and experience traces on specific tasks; they especially depend on extensive human engineering of reward functions, shifting the bottleneck from data curation to reward curation[86].

Frontier AI models have advanced by learning from high-quality task-solving trajectories, increasing inference-time computation, and incorporating human attempts; together, these factors have enabled advances such as solving the Navier-Stokes problem[91, 92].

Although some argue these advances are currently bought with inference-time compute rather than efficiency, it cannot be denied that human insight shrank the search space enough for brute force to finish OpenAI's Navier-Stokes resolution[93] in 88 hours[94].

{*These 88 hours are the run from the start of the effort*[93] *to its reported resolution*[95]*. The result establishes a forced Navier-Stokes singularity at every fixed viscosity, covering the Clay problem's breakdown statements (C) and (D), along with two Euler singularities. The unforced problem, statements (A) and (B), remains open, and any search for it first needs a target*[95, 96].}

This reported time did not include the formal verification, which took 17 more hours[94] of Lean work using GPT-6 Astra. The total time was therefore about 105 hours [88hrs for resolution + 17 hours for verification][94].

A close observation of the above reported results shows us the following

- High-profile results such as the claimed AI-generated Navier-Stokes proof exemplify resource-intensive progress
- This reportedly required around 10,000 coordination agents and roughly 130 billion output tokens[94, 97, 98]; it also showed the 'asymmetry' between costly generation and cheap verification: i.e. the search took about 88 hours, while formal checking in Lean took 17 hours[94, 97, 98].
- About 100 agents solved the unforced Euler problem in about 50 hours[94, 97, 98] (100 agents × 50 hours is roughly 5,000 agent-hours), under 1% of the roughly 10000 agents x 88 hours = 880,000 agent-hours of the main run[94, 97, 98].
- This shows that a swarm of agents two orders of magnitude smaller produced a reusable intermediate result, underscoring the value of "Decomposing and Reusing"[94, 97, 98], indicating that a frugal approach to problem decomposition helps beat brute-force search.

To summarise, Frontier AI still lacks an agreed definition. The reported Navier-Stokes solving effort showed that generating a result costs far more than verifying it, and that decomposing problems, guided by human insight, can outperform brute-force scale.

Frugal AI could leverage all these under

- A frugal research agenda could treat such discovery-scale compute as a one-time and disclosed cost, and thus can concentrate on distilling the strategies found into verifiable, accessible tools and thus measure the progress by cost per verified result rather than elapsed time for total computation [including solving and verification]
- Verifiable rewards, such as correctness checks, unit tests and proof checkers, are becoming cheap to compute and objective to assess. Under Responsible Frugal AI, they can provide the strongest evidence that efficiency gains preserve capability, establishing efficiency as a form of responsible engineering superiority.
- From a Responsible Frugal AI perspective, research workflows are likely to be most compute-efficient when they draw on genuine domain expertise. For example, domain specialists can frame problems well, break them into tractable sub-problems, recognise promising directions early and verify the results. This could narrow the search for solving hard problems that might otherwise require brute-force computation. This opens an interesting case for our nation, as it points to a strategy of investing in mathematicians and domain scientists and building human-centric AI that amplifies their expertise rather than replacing it with sheer scale.

- Technically, RLVR [Reinforcement Learning with Verifiable Rewards] rewards[80, 82, 83], when considered, for example, along with Lean/Mathlib[99, 100], could also replace expensive human labelling with cost-effective, deterministic checks. Human expertise could be leveraged for shared, reusable verification infrastructure rather than spending on expensive labelling outputs; i.e., without spending on frontier-scale labelling infrastructure, we can leverage RLVR that requires only modest reward-computation cost[79, 81].
    - This is a strong leverage point for Responsible Frugal AI that is building indigenous verifiers and benchmarks, libraries, developmental frameworks, etc for India, spawning agricultural, medical, legal, or any similar engineering tasks. They are a low-capex route to RL-driven capability on small, open, domain-specific models. Each verifier, benchmark or library built once can be reused by every model that follows. It turns domain expertise into a durable national asset, and it lets RL on small open models do the work instead of frontier systems.
- Responsible Frugal AI could account for the entire AI lifecycle by studying the effectiveness of their resources through extending the TCO beyond expenditure, revenue, and profit margins to energy, carbon, water and embodied hardware.
- Also, under Responsible Frugal AI, any amortisation analysis could be included, turning AI lifecycle total into a per-task figure.

TCO analysis, lifecycle accounting, amortisation, and per-task comparisons, thus considered in the Responsible Frugal AI framework, provide a comparative analysis of a distilled model against a teacher or against using an API or small domain-specific models [with RLVR] gaining from human expertise under a human-AI-centric ethical framework.

TCO, constraint-based requirements, a cap, and a safety margin, while also resisting rebound effects[101], provide a framework against which we could evaluate them; the associated evidence measures would show whether the returns of these advancements are sustainable and whether they suit India-specific cases. That is one of the important outcomes of Responsible Frugal AI

## v. Frugal AI: Why now?

Choosing problems, models, and systems should depend on whether further scaling, novel techniques, or reasoning research justify their financial, environmental, and institutional costs. This question remains ill-defined, hard to measure, and rarely asked upfront.

Frugal AI turns this from an afterthought into a design principle, demanding demonstrable value per unit of compute, energy, and external dependence before any scaling.

As a result, one of the most significant drivers of Responsible Frugal AI is turning a vague question into a design discipline by fixing the functionality a task requires, minimising resources and cost, and demanding robust, safe, and rigorous margins. Because rebound effects of efficiency gains can also raise total consumption, Frugal AI could limit total use to reduce resource demand[101].

Frugal AI measures innovation by fit to purpose, not just by model scale or just by compute consumption alone. When adding compute is easy and removing it is hard, the discipline to right-size models, data and infrastructure should be seen as the highest form of principled engineering. This means engineering restraint to deliver the most capability per watt, per rupee, and per byte is the most advanced AI skill of all.

Thus, the "Frontier of AI" need not be the largest model; it can also be the smallest or right-sized model that does the job, as envisaged by Frugal AI approaches.

# IV. India's AI Debate

India's AI debate is often pulled between multiple insufficient views

- One view assumes that national competitiveness requires replicating the most compute-intensive, capital-heavy foundation-model programmes developed in the United States or China.
- The second view is that India can remain a downstream adopter of foreign AI systems while adding localised interfaces and policy safeguards.
- The third view is related to the risks associated with AI. Given the AI threat landscape, model makers often reveal AI cyber threats that threaten the world order.
    - For the revealed threats, since they are universal, our nation can address them as the rest of the world would to mitigate the risks, as waiting to build from scratch might be too late. Our nation can do this while developing its own solutions.
    - But our nation faces other unique risks too, as reported by CeRAI in its AI Incident Reporting Framework for India[102];

The "first view" underestimates AI's material constraints: high-end GPUs, large-scale storage, memory bandwidth, advanced networking, energy supply, cooling, semiconductor dependence, and recurring model-training expenditure. These constraints are not just technical; they are fiscal and geopolitical. Public-interest AI for education, agriculture, healthcare, welfare delivery, judicial assistance, and language access cannot rely on assumptions of permanently high infrastructure costs.

The "second view" is also weak because downstream dependence limits India's room to shape model behaviour, retain control over sensitive data flows, and align AI systems with local languages, institutions, and public-service requirements. Reliance on imported model APIs and foreign cloud control may be acceptable in low-risk commercial applications, but it becomes structurally problematic in high-volume public systems and strategic sectors.

The "third view" calls for unique solutions from India, a hybrid pathway, and it raises a direct question of what is achievable (adoption vs indigenous). Unlike the first two views, the third view treats shared, time-sensitive threats as ones to be adopted and argues that adoption is the best rationale solution, while indigenous development to mitigate threats specific to our nation can be pursued accordingly.

Frugal AI approaches adopt this hybrid pathway. Frugal AI approaches and domain-specific models need not depend on frontier AI models [refer to section III, subsection iv] and can connect directly to the issue at hand.

Frugal AI in India is especially significant because its safety margin can include robust operation under unplanned conditions and recovery from failure, since AI model makers often attribute failure to AI cyber threats. This approach can draw on incident reporting in India, such as from CeRAI[102]. And Frugal AI, from an India-centric view, should also indulge in engineering new advanced frugal hardware made locally but adopted globally, thus supporting the Make-In-India initiative.

Accordingly, India could focus on threats unique to its sovereign boundaries rather than developing solutions from scratch for revealed threats. This is most cost-effective, and Frugal AI becomes especially significant.

Frugal AI, as shown above, offers a solid foundation for the third path while accounting for indispensable risks of AI adoption given the AI threat landscape, both global and unique to India. Frugal gains significance because it reuses globally validated defences against universal threats, concentrating scarce indigenous effort on India-specific risks that imported, general-purpose solutions cannot adequately address. This hybrid pathway stands out in this view because security is the only view where delay has an adversarial, imposed cost.

Also, Frugal AI questions are not just about “How can India build the largest possible model?” but “For any AI tool-based solution, what mix of data, models, hardware, and public infrastructure will maximise reliable social value per rupee and per joule?” That question and path are more relevant to India’s developmental reality.

The background rationale is that India's constraint set is distinctive, given unequal connectivity and device capability, budget-constrained state capacity, extreme linguistic and cultural diversity, and heavy reliance on public digital rails (Aadhaar, UPI, DigiLocker, ONDC, BHASHINI) for scale. High-end AI hardware is scarce and costly, and the binding bottleneck is computing access, not ambition.

These are not just gaps to bridge; they are design inputs. For example, Frugal AI treats compute, energy, data, and capital, along with risks and a safety margin, as first-class constraints[read design requirements and specifications] and maximises social impact per unit of energy spent [or a cost factor].

## V. Why India needs Frugal AI? Atma Nirbhar AI

India is a resource-imbalanced nation. It is neither resource-poor nor uniformly resource-constrained. Resources such as compute, connectivity, skills, and capital exist, but they are unevenly distributed across regions, languages and institutions.

Deploying public solutions—such as education and healthcare—requires unprecedented scale and is complex and India-specific. Thus, India needs to build its own sustainable AI ecosystem for its massive population and diverse regional coverage and penetration, addressing imbalances. For a nation of India's scale, complexity constraints, and diversity, any argument that presupposes computationally intensive foundation models as the default path, especially for its public system governance, including strategic and critical sectors, does not address India’s specific needs.

Three linked claims support this India-specific need.

1. **First**, Advanced Frugal principles are rigorous science, and their application in AI is the most viable path to large-scale deployment and operation under India’s fiscal, infrastructural, and computational on-the-ground realities. AFIs created using Indian frugal engineering principles with cutting-edge research in science and technology will result in high-quality but affordable products in all sectors.
2. **Second,** Advanced Frugal principles are a more credible route to AI sovereignty because they are inputs to design requirements and specifications rather than a retrofitted goal that is built on the most capital-intensive frontier model-building strategies and autonomy-primacy-focused agentic AI system solutions that are pursued elsewhere as a foundation.
3. **Third,** India's digital public infrastructure model aligns well with frugal AI, as it prioritises reusable public rails, modular architectures, local adaptation, and accountable deployment, operations and an infrastructure that is not built on any specifically imposed standards drafted elsewhere.

It follows that India needs its own AI – Atmanirbhar AI; i.e. India’s AI vision and goals should not imitate frontier-scale model races or Agentic AI compulsions, but they should rather orient around national needs and public purpose.

The strongest basis for this strategy is Frugal AI, with strong recommendations for India's relevant AI strategies as summarised below[103–107]

- A mature absorption stack lowers the capability threshold a model must meet, because the environment supplies the structure and verification the model would otherwise have to supply itself[106].
  - Crossing this absorption frontier could extract enough value from smaller, older or open-weight models to gain resilience against limited frontier access.

- India's digital public infrastructure has laid the first two layers, legibility and actionability, but verifiability, the hardest requirement, remains to be built[106].
  - The opportunity for Responsible Frugal AI is to leverage downstream value creation immediately, beyond downstream dependence[105].
  - This means deploying AI at scale on right-sized models and public rails that India controls.
- The (future) AI opportunity for our nation lies not in competing with large labs to build better models, but in finding the many ways AI can be deployed to create value[105].
- The GOI has announced through the IndiaAI Mission[107] support for Sovereign AI Infrastructure and Indigenous foundation models, and through the IndiaAI Compute Portal[104] enabling industry collaboration, ensuring socially impactful AI projects and bolstering ethical AI to drive responsible, inclusive growth of India's AI ecosystem, accelerating digital transformation and the economy.

# VI. Responsible Frugal AI from CeRAI

The GOI had already announced its strategy to achieve "Sovereign AI Infrastructure " through the India AI Mission and Semiconductor Initiatives[107] and India Compute Portal[104] and IndiaAI capacity[108]. They democratise access to AI compute resources at significantly lower costs, making compute available at affordable rates for Academia, Researchers, PhD Scholars, Students, Startups, MSMEs, and industry[103, 104, 108, 109].

Accordingly, CeRAI-IIT Madras proposes diffusing Frugal AI across a stack of four resources where Frugality can be engineered. These four resources are

- Data,
- Compute,
- Model, and
- Deployment and operational scenarios ( including use cases that justify them)

The following is the list of proposed Engineering Principles for Responsible Frugal AI, and thus frugality Levers across the AI Lifecycle

a) Data Foundation: What a model learns from, and how much of it.
b) Architecture Choice: What kind of model processes sequences.
c) Training Efficiency: How the forward and backward passes are made fast.
d) Compression: Making a trained model smaller and faster
e) Adaptation: Customising Frugally across languages, domains, and tasks
f) (Retrieval) Grounding: Adopting methods such as RAG[110] to reduce dependence on parametric memorisation by supplying knowledge at inference time from external, citable sources, which makes outputs attributable and checkable and lets smaller models match larger ones on knowledge tasks and also by reducing how much a model needs to memorise parametrically[111, 112].
g) Serving and Silicon: Including frugal engineering and frugal design choices that must be made, especially while deploying downstream, considering training efficiency, inference efficiency and deployment efficiency at every device tier[105, 106].

Recapping, the four resources (data, compute, model, and operations) are where frugal principles apply directly and can be viewed as "Four Frugality Dimensions". The seven levers are the practical means of making them. Most levers save mainly in one dimension but bring smaller savings in others too

## a. Data Frugality

Data frugality means using the least data that meets the requirement, and making that data clean, documented and well chosen. Data frugality does not mean using less data than a model needs. When requirements are demanding, or users are poorly served today, the data need can be large.

Indian language resources and highly specific regional datasets for local agriculture or health conditions remain severely limited. However, recent research demonstrates that small models trained on highly curated data can outperform massive models trained on unfiltered internet crawls[113]. Datasets like Fine Web-Edu further show that meticulously filtered text dramatically improves a model's reasoning capabilities, suggesting India can build powerful AI using compact, high-quality local data[114, 115].

Practices:

- De-duplicating and checking train–test overlap before training[116–118].
- Efficient use of data budget by active labelling and runtime curation of data and judicious use of synthetic data from domain knowledge.
- Documenting datasets and publishing their composition and intended use.
- Pruning or curating where the pruning cost is lower than the savings.
- Reusing scarce data within the range the evidence supports, and reporting where the returns decay.
- Report performance per language and per group, not as a single average.
- Measures
    - i. Dataset size and labelling effort per task
    - ii. Overlap rate between training and test data
    - iii. Per-group accuracy
    - iv. The share of data spent on underserved languages.

## b. Compute Frugality

Compute frugality means meeting the requirement with less training and inference compute and measuring that use objectively. It has an efficiency part (doing the same with less) and a reporting part (making the use visible).

In India, High-end AI hardware is scarce and prohibitively expensive, with a single NVIDIA H100 GPU costing up to $40,000, or roughly ₹30-40 lakhs in India[119] as of Nov 2026, and these numbers keep moving rapidly.

The Economic Survey 2025-26 explicitly warns that India's AI rollout is bottlenecked primarily by this global GPU shortage, not a lack of domestic funding[120, 121]. Because many rural regions lack reliable internet, deploying smaller, efficient models directly on local edge devices bypasses expensive cloud compute and ensures consistent access.

Practices:

- Match model size and training data to the compute budget.
    - a. Use compute-optimal scaling as the starting point.
        - i. Given a fixed training compute budget, how should it be split between model size and training data so that the loss is as low as possible?
    - b. When a model serves a high volume of queries, train a smaller model on more data than the compute-optimal point
    - c. This suggests that the extra training cost can be compensated through cheaper (read: lower) inference costs.

- Use lower-precision arithmetic, such as mixed-precision training and 8-bit or lower-precision inference, wherever the model still meets its declared performance threshold on targeted tests.
- Tune hyperparameters on smaller proxy models and transfer the best settings to the full-size model, instead of running full-scale sweeps (i.e. repeated training runs that test many combinations of settings).
  a. Confirm that the settings transfer with a short check at the target scale.
- Build on existing work rather than training from scratch.
  a. Start from available checkpoints (saved snapshots of a model during training) or open-weight models (models whose trained parameters are publicly released) and adapt them to the task.
  b. Publish the checkpoints so that other teams can build on them instead of repeating the same training.
- Control the compute a model uses when answering queries /or when in use (inference).
  a. Set a budget for how many reasoning steps and repeated attempts a model may use per query, scaled to the difficulty of the task, so simple queries do not consume the effort meant for hard ones.
  b. Avoid repeated and wasted computation when serving queries. Store and reuse results for questions, or common parts of questions, that recur frequently (caching), instead of computing them afresh each time.
  c. Group incoming queries and process them together (batching), so the hardware runs at fuller capacity, and each query costs less, within acceptable response times.
- Reporting energy (kWh) and carbon ($tCO_2e$), including embodied emissions, for every training run and for deployment and operation.
- Count all compute, including failed runs, ablations and hyperparameter sweeps, not just the final run.
- Using tracking tools and standardised appendices so that reports can be compared.
- Place and schedule compute where cleaner power is available.
- Having energy figures checked independently.
- Measures
  i. Energy (and carbon) per unit of useful outcome;
  ii. Sharing of embodied emissions;
  iii. The gap between reported and independently measured use.
  iv. Total GPU-hours, including failed and exploratory runs
  v. Hardware utilisation (share of available compute doing useful work)
  vi. Inference compute per query or per completed task

### c. Model Frugality

Model frugality means choosing the smallest model, or the smallest set of trained parameters, that meets the requirement and verifying it against that requirement. In CeRAI's formulation, model frugality is entailed by a systematic search for alternative designs that yield resource savings. This is evaluated along the Pareto front of task performance, computational cost and memory footprint. Every model design involves trade-offs. Better performance usually needs more compute or memory. The Pareto front is the set of designs that make these trade-offs efficiently. The preferred design is not the one that maximises performance, but the point on the Pareto front that satisfies the declared performance requirement at minimum computational and memory cost. This means searching for an alternative design that delivers savings based on rigorous scientific and (frugal) principles. (Refer to Figure 1)

Also, it should be noted that agent harnesses and agentic workflows[122–126] could be a frugal route to capability because instead of sending every step to a frontier model, an orchestration layer could decompose tasks, use tools and caches, and route routine, narrow steps to small, fine-tuned models, escalating to a large model only for hard reasoning.

This heterogeneous design cuts cost, latency and energy per task, and keeps data on infrastructure the user controls[122–126].

In India, many standard tasks, such as crop disease classification or educational recommendations, do not require the massive computational overhead of generative AI. By using classic machine learning for core logic and reserving generative models strictly for translating outputs into local dialects, developers can "right-size" their applications. A rural agricultural advisory platform illustrates this hybrid architecture well: it can use a traditional, low-compute decision tree to accurately diagnose a crop disease from structured soil data, while a small, locally hosted language model translates that diagnostic result into a fluent, conversational Tamil interface. This targeted workflow lowers inference costs and makes the system much easier to guardrail against hallucinations.

Practices:

- Starting from the smallest candidate model and justifying any step up with test evidence.
- Using adaptation and compression before training from scratch.
- Evaluating with targeted tests of the task, not with general impressions.
- Reserving larger models for tasks that demonstrably need them
- Measures
  - i. Parameters and memory per task;
  - ii. Accuracy on targeted tests with uncertainty;
  - iii. The share of requests escalated to a larger model.

## d. Operational Frugality

Operational frugality means serving at low cost per useful outcome and sizing the model to each task, routing queries sensibly, using hardware well, and measuring resource use where an AI product, service, or resource is consumed, applied, or accessed. It matters because AI serving costs accumulate over the life of a system.

To operationalise this frugality at scale, India should build a public AI stack as Digital Public Infrastructure (DPI)—along the lines of the India Stack for UIDAI. It should be a layered, interoperable set of shared rails that any state department, startup, nonprofit, or public institution can plug into. At the base, this stack should include public datasets, consented data exchanges, Indian language benchmarks, model evaluation sandboxes, secure inference gateways, and open APIs for speech, translation, retrieval, classification, and lightweight generation. Above that, India should support a tiered model ecosystem: small local models for edge deployment, task-specific discriminative models for high-volume public-service workflows, and larger generative models only where they add real value in communication or reasoning. A DPI-style AI stack reduces duplication, lowers entry barriers, improves accountability, and makes it possible to deliver trustworthy AI as a public utility rather than as a collection of costly, siloed products. It should be designed so that public money funds reusable rails, shared safety standards, and curated domain assets, not repeated reinvention of the same models. India does not need a single sovereign supermodel; it needs sovereign public AI rails on which many right-sized models can run.

Practices:

- Treat AI as 'Infrastructure ': Treating AI as infrastructure means providing shared capacity (compute, data, evaluation, and serving) as public rails (such as DPI) that many small services can use, so each service need not build its own. CeRAI's proposition is that pooling spreads fixed costs across many users and lowers resource use per service.

- Building portable layers for data, evaluation, and serving
- Tying access to a stated requirement and reporting utilisation of the shared capacity.
- Keeping non-AI fallbacks and human handover for essential services.
- Measures:
  - i. Utilisation and idle capacity
  - ii. Cost per user served;
  - iii. Share of services using a non-AI fallback;
  - iv. Incident rates.

## VII. CeRAI's principles for Frugal AI

Frugality in AI should be understood as adequacy without waste. It does not always mean inferior or cheaper AI systems. Rather, it demands disciplined and rigorous frugal engineering: selecting architectures, datasets, inference pathways, interfaces, and governance mechanisms that fit the purpose and match the task. The following are the Frugal AI principles CeRAI has considered.

1. Ask first whether AI is needed. Begin with the problem. Use a rule, a decision guideline, or a form before an AI model. Deploy AI where it demonstrably beats the cheaper alternative.
2. Choose the smallest sufficient model. Declare a performance bar, then choose the smallest model that clears it. With this approach, use generative models only for tasks that need generation. Given this, pruning, quantisation, distillation and retrieval will become the default toolkit in the Frugal AI framework.
3. Frugality is the path to sovereignty. Consider this: a portfolio of open models [weights plus training data, code, and recipes] that are domain-specific Indic models (or task-specific SLMS) hosted on Indian soil for critical domains is a sovereign capability. Sovereignty means the ability to exercise responsible autonomy, specify design. deploy, govern, run, inspect, adapt and switch — not just the ownership of the largest clustered data. "Frugality in AI/Frugal AI should be understood as context-specific excellence. Gain in performance comes from deep knowledge of the task, the domain, the users and the operating conditions, not merely from dumping data, compute or large model size
4. Build AI as Digital Public Infrastructure. Public money funds reusable rails — shared datasets, benchmarks, safety tests, and open APIs for speech, translation, retrieval and lightweight generation — once, for everyone, rather than the same model many times over.
5. Design for the last mile first. Voice-first and vernacular-first by default. The first-time smartphone user, the intermittent connection and the low-end device are the reference case, not the edge case.
6. Treat diversity as a design objective, not an afterthought. Languages, dialects, regions and user groups are requirements to test against, with performance reported per group rather than as a single average.
7. Make resource use observable. What is not measured is not governed; hence, log and publish compute, memory, energy, water, data, and rupees per unit of outcome, alongside accuracy.
8. Keep humans in the loop where stakes are high. Teacher-in-the-loop, doctor-in-the-loop and officer-in-the-loop by default for consequential decisions, with audit and grievance routes built in.
9. Cap the total, not only the unit. Cheaper inference invites more inference. Every frugal programme declares an absolute resource ceiling and reports against it; efficiency gains that are spent on volume are not frugality.

# VIII. CeRAI's Proposal: Responsible Frugal AI Framework

CeRAI's Responsible Frugal AI framework inverts the "Frontier by default" approach: it starts from what a computational governance task in India requires and treats any additional scale as a cost to be justified. (Refer to Section III).

The following are other significant factors to consider for the Responsible Frugal AI approach

- Lightweight capabilities and Frugal approaches to be considered default solutions before any frontier-model procurement sign-off. We need to ask this: "Does this task genuinely require frontier-only capability, or does a domain-tuned small/medium model suffice or no AI at all?
- India already has the foundations of a sovereign model portfolio, and efficiency techniques such as compression and quantisation are well known. However, systematic, widespread use to build domain-specific models sized to the task is needed. Addressing any shortfall during implementation requires the Frugal AI framework, because Frugal AI approaches include explicit principles, measures, and adoption criteria.
- With the Frugal AI approach, we could see how to track the counterfactual claims explicitly: For example, if we logged every instance where a frontier model was used but a smaller sovereign model or Frugal approach would have actually sufficed, we could quantify a dependency cost (compute spend, data exposure, energy footprint, governance exposure, etc.). This metric could help us develop the right fit policies in the AI era.

These factors operate under the principles in Section VII. The policy commitments that follow are set out in the Policy Statements, and the deployment criteria are in the Adoption Checklist.

# IX. A Seven-Point Adoption Checklist

A Frugal AI deployment is ready when these minimum checklist items are ready

- AI need has to be established and is used only where it materially improves public value or workflow performance over a rule, a decision guideline or a form.
- The constraints such as power, bandwidth, device class, capex/opex, carbon budget, and regulatory limits shall be declared first and written down before any model is chosen.
- The team can show why the smallest sufficient model was selected, or a smaller, specialised, or composable model was rejected in favour of the chosen model — (and not the reverse). Rules, retrieval and classical ML shall be counted.
- The system shall be deployment-condition tested and should be evaluated under real Indian bandwidth, device, language, energy and institutional constraints.
- Resource use should be instrumented, i.e., compute, memory, latency, energy, lifecycle cost and rupees are to be logged and reported per unit of the outcome, along with accuracy and thus audited across the model/product lifecycle starting from training, fine-tuning, testing, deployment to estimated lifetime use. This implies that we can't judge whether an AI system is frugal at a single point in time; we have to judge it across the system's whole life: design, data collection, training, deployment, operation, maintenance, retraining and retirement.
- TCO and sovereignty are checked before scale. Total cost of ownership (GPU leases, bandwidth, vendor lock-in, retraining) and data jurisdiction are assessed before any population-scale rollout.
- Governable. Documentation, auditability, non-AI fallback paths and human oversight are in place for consequential use.

# X. Responsible Frugal AI from CeRAI – The Policy Statements

AI is currently shaped by a default that the largest model is the right starting point. That default is expensive in energy, capital and institutional capability, and its costs are not distributed equally. For India, digital public services and computational governance tasks must reach very large, diverse populations at a rationalised cost.

CeRAI's Frugal AI white paper makes Frugal AI part of its commitment and mandate for Responsible AI by fixing the functionality that digital public services and computational governance tasks require, minimising resources and cost, and maintaining the rigour (a verified margin and robustness) that stops a lean design from failing.

This also gives developers, funders, and regulators something verifiable and evidenced; for example: Did the system meet the stated requirement, with evidence, and at what Total Cost of Ownership?

Efficiency techniques alone do not deliver frugality. The paper therefore combines efficiency techniques with a reviewable limit on total use for publicly funded AI.

1. Frugality should be treated as the default engineering principle for public-interest AI in India.
2. India should prioritise right-sized, domain-specific, and multilingual AI systems over prestige-driven model maximalism.
3. Public investment should build shared AI rails, not isolated, duplicative stacks.
4. Sovereignty should be pursued through modularity, local control, interoperability, and Institution-building, preferring smaller, indigenously adaptable AI systems that India can design, deploy, operate, inspect and maintain.
5. Diversity should be treated as a deployment requirement, not an afterthought.
6. Generative AI should be used selectively, where generation is essential to the task.
7. Public AI systems should be evaluated on utility per unit resource, deployment realism, accountability and public value, not on raw capability or benchmark success.
8. State capacity for Frugal AI should be distributed across ministries, states, universities, and public institutions.
9. India's AI future should be built around scalable public purpose, not imported assumptions of infinite compute.
10. The principles of advanced frugal innovation should be aligned with the India AI Governance Guidelines (2025) and supported by published implementation guidance that operationalises the stack-based approach.
11. Incidents and failures in Frugal AI deployments should be published, so that lessons can be learned

India does not need an AI strategy defined by imitation of the most capital-intensive global model race. It needs an AI strategy aligned with its own structural realities: public scale, uneven infrastructure, linguistic plurality, fiscal constraints, and the strategic need for institutional control. Advanced frugal innovations and the discipline of frugal engineering, with their genesis in India, offer a very potent mix for developing local Frugal AI that can plausibly win global appeal. The Frugal AI Framework for India from CeRAI operationalises this through its proposal of a Responsible Frugal AI approach for India.